# LIQUID-GAS PHASE TRANSITION IN HOT NUCLEI: CORRELATION BETWEEN DYNAMICAL AND THERMODYNAMICAL SIGNALS

M. F. Rivet,[1,][*] B. Borderie,[1] R. Bougault,[2] P. Désesquelles,[1] E. Galichet,[1,3]
B. Guiot,[4,][†] N. Le Neindre,[2] M. Pârlog,[5] G. Tăbăcaru,[5,][‡] and J.P. Wieleczko[4]
(for the INDRA collaboration)
[1]Institut de Physique Nucléaire, IN2P3-CNRS, F-91406 Orsay Cedex, France.
[2]LPC, IN2P3-CNRS, ISMRA et Université, F-14050 Caen Cedex, France.
[3]Conservatoire National des Arts et Métiers, F-75141 Paris Cedex 03.
[4]GANIL, CEA et IN2P3-CNRS, B.P. 5027, F-14076 Caen Cedex, France.
[5]Nat. Inst. for Physics and Nuclear Engineering, Bucharest-Măgurele, Romania.

*Abstract*

The dynamics and thermodynamics of phase transition in hot nuclei are studied through experimental results on multifragmentation of heavy systems (A$\geq$ 200) formed in central heavy ion collisions. Different signals indicative of a phase transition studied in the INDRA collaboration are presented and their consistency is stressed.

## 1. INTRODUCTION

Liquid-gas type phase transitions are commonly observed in systems with short-range repulsive and longer-range attractive forces, such as macroscopic fluids with the van der Waals interaction. The similarity of the equations of state for non-ideal gases and nuclear matter allows to foresee the existence of different phases of nuclear matter. Now can one define phase transitions for nuclei, small finite objects? Recent works state that statistical mechanics based on the Boltzmann's definition of entropy allows to define phase transitions in small systems [1]. The nucleus at zero or moderate temperature, because of its quantal nature, is assimilated to a liquid phase. A nuclear gas phase was characterized, for instance, through the properties of vaporised quasi-projectiles from 95 A.MeV Ar+Ni reactions, fully predicted by modelling a van der Waals gas of fermions and bosons in thermal and chemical equilibrium [2]. The liquid-gas coexistence region was thus naturally connected to multifragmentation, break-up of a nuclear system in several fragments of various sizes.

Many signals evidencing phase transitions were recently proposed in the literature [3–7]. The INDRA collaboration took advantage of the large body of data acquired so far to evidence and cross-check several of these signals. This paper presents results on negative heat capacity and spinodal decomposition obtained for events from *central collisions* between heavy ions ($A_{proj}$ + $A_{tgt}$ > 200), insisting on the correlated observation of these two signals.

## 2. PHASE TRANSITION: DYNAMICS AND THERMODYNAMICS

In statistical physics of finite systems, interface effects are not negligible with respect to bulk properties. This induces, in the coexistence region, anomalous curvature in thermodynamical potentials [8], with the direct consequence that both the heat capacity and the compressibility are negative [9]; this indicates that the system is in the spinodal region of the nucleus phase diagram. These considerations show the route to the search for signals of phase transition; one is related to the thermodynamical aspect, and consists in evidencing a negative microcanonical heat capacity. Another one is to look for the mechanism connected with the dynamics of the transition (which is a finite time phenomenon).

### 2.1 *Charge correlations and spinodal decomposition*

In nuclear systems, two mechanisms are envisaged for the dynamics of first order phase transition, nucleation and spinodal decomposition. The latter was widely investigated in theoretical studies and found responsible for multifragmentation in central heavy-ion collisions around the Fermi energy [10, and references therein]. Spinodal decomposition presents the interesting property of ending up with equal-sized fragments among smaller (gas) products. The clean signal expected in ideal infinite nuclear matter, where the size of the fragments is connected to the wave length of the most unstable mode, is however blurred in finite nuclei [11]; the remembrance of the spinodal process is thus expected to be weak in real nuclear systems. Powerful tools such as intra-event correlations must thus be used and memory of spinodal decomposition was searched for in scrutinizing the partitions resulting from multifragmentation, through charge correlations defined as $C = Y_{cor}(\langle Z \rangle, \sigma_Z)/Y_{uncor}(\langle Z \rangle, \sigma_Z)$ where $\langle Z \rangle$ and $\sigma_Z$ are the average charge and the standard deviation of the fragment ($Z \geq 5$) charge distribution inside each event. Spinodal decomposition should favour partitions with equal size fragments or small $\sigma_Z$, which will appear as peaks in the correlation function.

---

[*]e-mail:rivet@ipno.in2p3.fr
[†]Present address: Sezione di Bologna and Dipartimento di Fisica, Università di Bologna, Italy.
[‡]Present address: Cyclotron Institute, Texas A&M University, College Station, Texas 77845, USA.

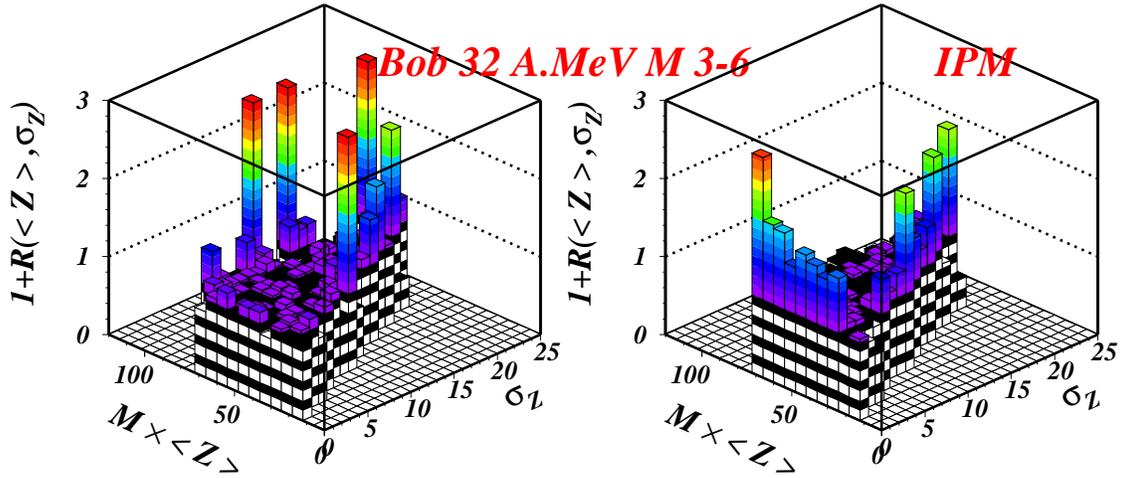

Fig. 1: Charge correlation function as a function of the fragment multiplicity times the average charge and of the standard deviation, for events generated by stochastic mean field simulations of central 32 A.MeV Xe+Sn collisions, built with EMM (left) and IPM (right).

Building correlation functions is never an easy task, because there is no unique way of evaluating the denominator. To evidence small signals, it appeared necessary to improve even more the initial method of building higher order charge correlation functions [12], by introducing the constraint of charge conservation. Two methods were recently developed in this aim: i) an event mixing method (EMM), where the uncorrelated events are built by exchanging two pairs of fragments with the same total charge between two events [13]; ii) an intrinsic probability method (IPM) which takes into account, in a combinatorial way, trivial physical correlations such as those due to total charge conservation [7, 14].

It appeared instructive to confront these methods to a sample of events obtained in a complete stochastic mean field simulation of central collisions between Xe and Sn at 32 A.MeV (the Brownian one-body dynamics - Bob): in the course of the collision, the system is driven into the spinodal region, *where it multifragments through spinodal decomposition* [15]. The de-excitation of the hot fragments so formed is followed simultaneously with their propagation in their mutual Coulomb field. Finally the simulated events (Bob events) are passed through a software replica of the INDRA detector. Numerous experimental features of the collisions were succesfully accounted for by this simulation [16].

The charge correlation functions built with the two methods are presented in fig 1, and yield contradictory results: a significant enhancement of partitions with small $\sigma_Z$ is visible with IPM, which is not observed with EMM. Remind at that point that spinodal decomposition *has occurred* in the collisions sampled here. From the EMM picture, one may conclude that either there is no discernable memory of the process, or that the method is not sensitive enough to evidence it, at least for the present size of the event sample ($4 \times 10^4$ events); thus EMM, in this case, does not furnish any information on the occurrence of spinodal decomposition. Conversely the IPM pleads in favour of a measurable memory of the process, and evidences it. The percentage of extra-events corresponding to $\sigma_Z < 1$ is indeed small (~0.4% of the total number).

## 2.2 Heat capacity

The microcanonical heat capacity can be derived from the fluctuations of the kinetic part of the total energy. It turns negative when these fluctuations overcome the reference (canonical) value [17]. This signal proved more robust than an eventual backbending in the caloric curve, the existence of which depends on the path followed by the system in the temperature-pressure-energy space [18]. The proposed derivation of the heat

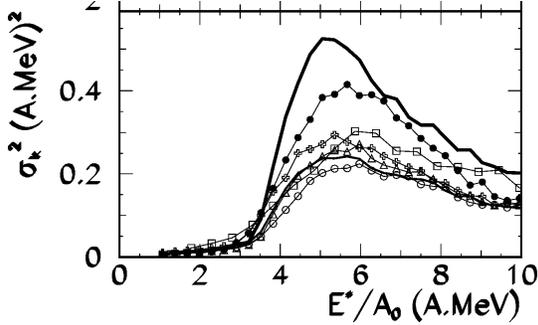

Fig. 2: Kinetic energy fluctuations for SMM simulated events. The true fluctuations in the model are represented by the solid line, and those 'measured' by using all the assumptions needed in experiments (including detector filter) are displayed with squares. From [19]

capacity must be done for the freeze-out configuration, which has thus to be reconstructed from the final one; the reconstruction, in the case of experimental data, requires various assumptions on the unmeasured quantities (neutrons, fragment masses, percentages of evaporated particles ...). Fig. 2 shows, for SMM simulated events, that these assumptions do not enhance but rather underestimate the fluctuation - provided that unknown quantities are replaced by averages - and thus do not artificially create a signal [19].

The above considerations show that if multifragmentation occurs in the spinodal region, one may observe simultaneously a negative heat capacity and a signal of spinodal decomposition. It is worth noticing that while any evidence of spinodal decomposition *must* be accompanied by a negative heat capacity, the reverse is not true, because the dynamical mechanism of the phase transition would be nucleation for instance, or because the spinodal instabilities would not have time to develop, due to a faster collision time.

## 3. EXPERIMENTAL INDICATIONS OF MULTIFRAGMENTATION IN THE SPINODAL REGION

Two reactions were studied which lead to systems with close total mass and charge ($Z \sim 105$, $A \sim 250$). In both cases samples of events resulting from central collisions between 30 and 50 A.MeV were characterised as multifragmentation of "fused" systems. In this paper we shall present results for the symmetric $^{129}$Xe+$^{nat}$Sn system at 32, 39, 45 and 50 A.MeV and for the asymmetric $^{58}$Ni+$^{197}$Au system at 32 and 52 A.MeV. Data concerning heat capacity and charge correlations, obtained on the same samples in all cases, will be presented in the first subsection, before being globally analysed in the next subsection.

### 3.1 Experimental results

Charge correlation functions obtained with IPM are displayed in fig. 3 for Xe+Sn at four incident energies; events with 3 to 6 fragments are summed in these plots. Significant enhancement of events above the background are observed for small values of $\sigma_Z$ [7]. The total number of extra-events for $\sigma_Z < 1$ is of the order of 0.1-0.3%, depending on the energy. These percentages are similar to those observed in the Bob simulations, and may be taken as an evidence of the occurrence of spinodal decomposition in central collisions between Xe and Sn. As for simulated events, note that charge correlation functions for all experimental samples presented here and built with EMM do not indicate significant enhancement of any type of partitions [13], The numbers of events in the samples are similar in all cases (3-$4\times10^4$ events); the results from this method will not be further discussed in this paper as it was stated precedently that in the present conditions they do not furnish information on the occurrence of spinodal decomposition.

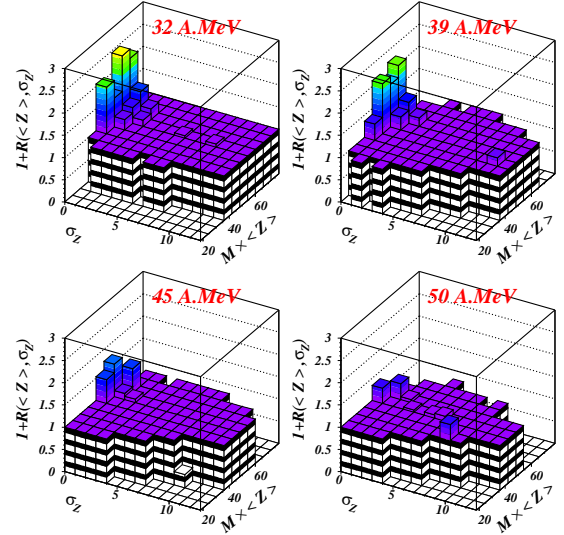

Fig. 3: Charge correlation functions for central collision events between Xe and Sn at 4 energies obtained with the IPM. Peaks (holes) corresponding to a confidence level lower that 90% have been flattened out. From [7].

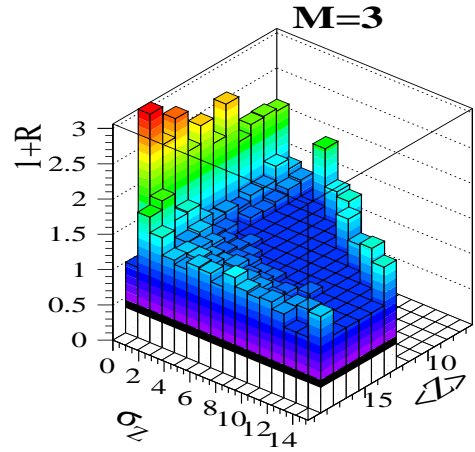

Fig. 5: Charge correlation function for central collisions producing 3 fragments between Ni and Au at 52 A.MeV, with IPM. From [20].

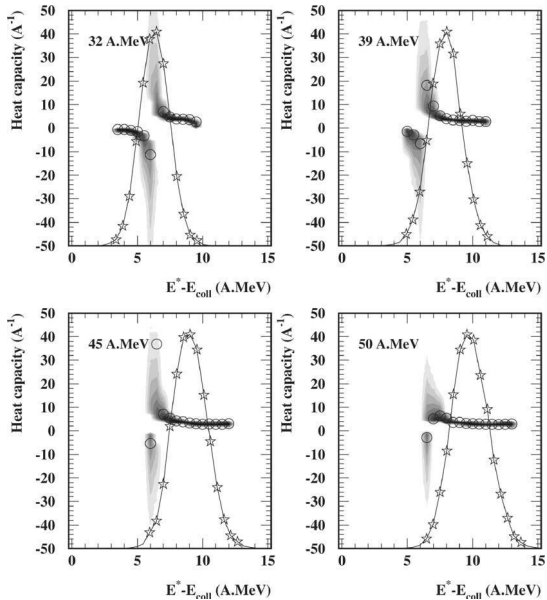

Fig. 4: Excitation energy distributions (full lines) and heat capacity (shaded zones) measured in central collisions between Xe and Sn. From [19].

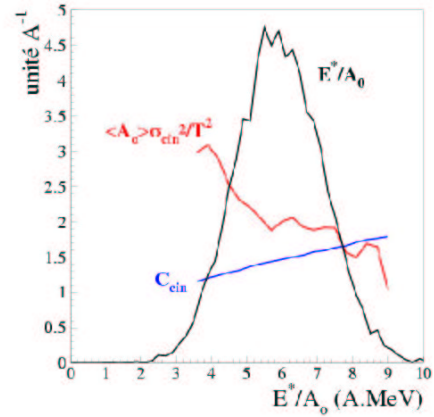

Fig. 6: Excitation energy distribution, kinetic energy fluctuations and reference heat capacity for central Ni+Au collisions at 52 A.MeV. From [20].

The heat capacity for the Xe+Sn systems is shown in fig. 4. The excitation energy scale was obtained by calorimetry, from which was subtracted a radial expansion energy, determined from a comparison of the data with results of a statistical multifragmentation model. Negative branches are observed below $E^*$ =6-6.5 A.MeV at the four incident energies. At 32 and 39 A.MeV there is a significant overlap between the energy distribution and the negative heat capacity; this indicates that multifragmentation occurred in the spinodal region, in agreement with the signal given by the charge correlations.

For the asymmetric entrance channel Ni+Au, a significant number of enhanced partitions of equal-sized fragments was observed, with the IPM, at 52 A.MeV for events with 3 (fig. 5) and 4 fragments. No significant signal was seen at 32 A.MeV. The heat capacity calculated at 52 A.MeV also shows a negative branch, as shown in fig. 6 where the kinetic energy fluctu-
ations overcome the reference heat capacity for excitation energies between 4 and 8 A.MeV. The heat capacity remains positive at 32 A.MeV when determined on the same sample as the charge correlations [21]. Note that a negative heat capacity was observed for Ni+Au at 52 but also at 32 A.MeV on differently selected samples of central collision events [22]; this is related to detection thresholds which render rather poor the completeness of the samples in the case of Ni+Au at 32 A.MeV (about 65% for the total detected charge, to be compared to more than 80% required for Xe+Sn). For this asymmetric entrance channel, multifragmentation in the spinodal region is reliably shown by dynamical and thermodynamical signals at 52 A.MeV

### 3.2 *Synthesis of the observations*

Table I: Summary of the phase transition signals described in the previous section. $\varepsilon_{inc}$, $\varepsilon_{th}$, $\varepsilon_{rad}$, stand for the incident, thermal and expansion energies in A.MeV (see text for precisions). SD means signal of spinodal decomposition.

| system | Ni+Au | Xe+Sn | Xe+Sn | Ni+Au | Xe+Sn | Xe+Sn |
|---|---|---|---|---|---|---|
| $\varepsilon_{inc}$ | 32. | 32. | 39. | 52. | 45. | 50. |
| $\varepsilon_{th}$ | 5.0 | 5.0 | 6.0 | 6.5 | 6.5 | 7.0 |
| $\varepsilon_{rad}$ | 0. | 0.5 | 1.0 | 0. | 1.5 | 2.2 |
| SD | no | yes | yes | yes | yes | no |
| c<0 | no | yes | yes | yes | no | no |

A synthesis of the signals concerning spinodal decomposition and heat capacity for nuclei of mass ∼250 is presented in table I. The thermal and radial energy scales come from SMM simulations, with respective uncertainties of 1 and 0.5 A.MeV, depending, for instance, on the event selections, the freeze-out volume chosen .... The first evidence coming from this table is that for the asymmetric entrance channel, multifragmentation of the fused system formed in central collision is *not* associated to a radial expansion. This fact is directly visible from the data: for Xe+Sn at 50 A.MeV and Ni+Au at 90 A.MeV (the latter data are not discussed here), the charge and fragment multiplicity distributions are identical, while the average kinetic energies of fragments are much smaller (by ∼ 20 MeV for Z=10-30) in the case of Ni+Au [23]; these observations indicate that the thermal energy, which governs partitions, is the same in the two cases while the expansion energy is small, or not existing, for Ni+Au. This comes at variance of other results for similar systems in the same energy range, for instance Kr+Au [24], where a significant expansion energy was reported.. From the table, it appears that, for both entrance channels, the systems reach the spinodal region and remain there for a time sufficient to allow the development of spinodal instabilities when the thermal energies lies between 5 and 6.5 A.MeV, as attested by the observation of favoured equal-sized fragment partitions and the negative values of the heat capacities. Radial energies up to 1 A.MeV do not influence the scenario. When the thermal energy increases, but also and relatively more the expansion energy, the system crosses too fast the spinodal region, toward the coexistence region or even the gas region. In this case neither the spinodal decomposition nor the negative heat capacity are observed. One can get more details in looking at the excitation function for extra-events, shown in fig. 7. Here events with $\sigma_Z$ <3 were considered (see fig. 3). One observes a rise and fall of the percentage of par-titions remembering the original process, with a maximum at 39 A.MeV, and no more memory at 50 A.MeV. The favoured production corresponds to $\varepsilon_{th}$ around 5.5-6 A.MeV, with a gentle compression around 1 A.MeV.

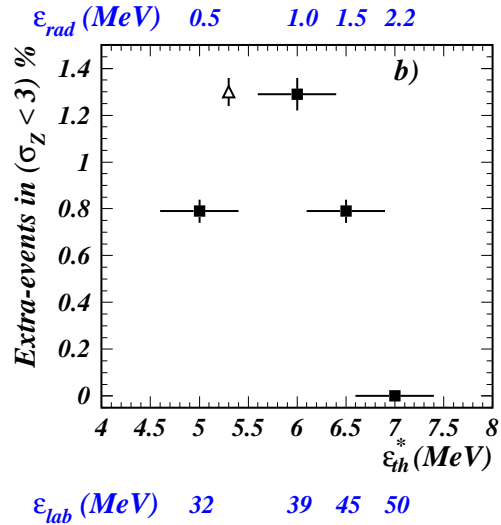

Fig. 7: Excitation function of extra-events with nearly equal-sized fragments versus the excitation energy, for Xe+Sn. The open symbol refer to the simulated 32 A.MeV sample. See table 1 for the energy scales and [7] for details.

## 4. CONCLUSIONS

The results obtained in analysing central collisions leading to A∼250 systems through a symmetric and an asymmetric entrance channel show several features which may be taken as characteristics of a liquid-gas phase transition in nuclei. While each of the two signals discussed in this paper may not be conclusive by itself, their concomitance gives strength to the assumed scenario. Other observed variables or signals characteristic of a phase transition were observed (see contributions of R. Bougault, B. Tamain, J. D. Frankland).

## 5. REFERENCES

[1] D. H. E. Gross, *Microcanonical Thermodynamics - Phase Transitions in "small" systems* (World Scientific, Singapore) (2001).
[2] B. Borderie *et al.* (INDRA collaboration), Eur. Phys. J. A 6 (1999), 197.
[3] M. D'Agostino *et al.*, Phys. Lett. B 473 (2000), 219.
[4] R. Botet *et al.*, Phys. Rev. Lett. 86 (2001), 3514.
[5] P. Chomaz *et al.*, Phys. Rev. E 64 (2001), 046114.
[6] J. B. Elliott *et al.*, Phys. Rev. Lett. 88 (2002), 042701.
[7] G. Tăbăcaru *et al.*, Eur. Phys. J. A 18 (2003), 103.
[8] D. H. E. Gross, T. Dauxois *et al.*, eds., *Dynamics and Thermodynamics of systems with long range interactions* (Springer-Verlag, Heidelberg), vol. 602 of *Lecture Notes in Physics*, 23–44 (2002).
[9] P. Chomaz *et al.*, T. Dauxois *et al.*, eds., *Dynamics and Thermodynamics of systems with long range interactions* (Springer-Verlag, Heidelberg), vol. 602 of *Lecture Notes in Physics*, 68–129 (2002).
[10] P. Chomaz *et al.*, Phys. Rep. 389 (2004), 263.
[11] M. Colonna *et al.*, Nucl. Phys. A 613 (1997), 165.
[12] L. G. Moretto *et al.*, Phys. Rev. Lett. 77 (1996), 2634.
[13] J. L. Charvet *et al.*, Nucl. Phys. A 730 (2004), 431.
[14] P. Désesquelles, Phys. Rev. C 65 (2002), 034604.
[15] A. Guarnera *et al.*, Phys. Lett. B 373 (1996), 267.
[16] J. D. Frankland *et al.* (INDRA collaboration), Nucl. Phys. A 689 (2001), 940.
[17] P. Chomaz *et al.*, Nucl. Phys. A 647 (1999), 153.
[18] P. Chomaz *et al.*, Phys. Rev. Lett. 85 (2000), 3587.
[19] M. D'Agostino *et al.*, Nucl. Phys. A 699 (2002), 795.
[20] B. Guiot, thèse de doctorat, Université de Caen (2002), tel-0003753.
[21] B. Guiot *et al.* (INDRA collaboration) Private communication.
[22] O. Lopez *et al.* (INDRA collaboration), G. Agnello *et al.*, eds., *Proc. Int. Workshop on Multifragmentation 2001, Catania, Italy* (2001), 21–26.
[23] N. Bellaize *et al.* (INDRA collaboration), Nucl. Phys. A 709 (2002), 367.
[24] C. Williams *et al.*, Phys. Rev. C 55 (1997), 2132.